\documentclass[twocolumn,PRE]{revtex4}

\usepackage[english]{babel}

\usepackage{graphicx}
\usepackage{amsmath}
\usepackage{amssymb}
\usepackage{wasysym}   

\usepackage{microtype}

\newcommand{\beq}{\begin{equation}}
\newcommand{\eeq}{\end{equation}}

\newcommand{\rmi}{\mathrm{i}}
\newcommand{\rme}{\mathrm{e}}
\newcommand{\vect}[1]{\mathbf{#1}}

\usepackage{xcolor}

\begin{document}

\title{Theta-point polymers in the plane and Schramm-Loewner evolution}
\author{M. Gherardi}
\email{marco.gherardi@mi.infn.it}
\affiliation{Universit\`a degli Studi di Milano, Dip.\ Fisica, Via Celoria 16, 20133 Milano, Italy}
\affiliation{I.N.F.N. Milano, Via Celoria 16, 20133 Milano, Italy}

\begin{abstract}
We study the connection between polymers
at the $\theta$ temperature on the lattice and Schramm-Loewner chains
with constant step length in the continuum. 
The latter realize a useful algorithm
for the exact sampling of tricritical polymers, where finite-chain
effects are excluded.
The driving function computed from the lattice model via 
a radial implementation of the zipper method
is shown to converge to Brownian motion of diffusivity $\kappa=6$
for large times. The distribution function of an internal
portion of walk is well approximated by that obtained
from Schramm-Loewner chains.
The exponent of the correlation length $\nu$
and the leading correction-to-scaling exponent $\Delta_1$
measured in the continuum are compatible with
$\nu=4/7$ (predicted for the $\theta$ point)
and $\Delta_1=72/91$ (predicted for percolation).
Finally, we compute the shape factor and the asphericity
of the chains, finding surprising accord with
the $\theta$-point end-to-end values.
\end{abstract}

\maketitle

\section{Introduction}

Self-avoiding walks and percolation are the quintessential \emph{geometric} critical phenomena.
In such systems, either the statistical ensemble or the properties investigated are of a geometric nature.
For instance, a self-avoiding walk (SAW) is a nearest-neighbor walk on a lattice
such that each site is visited at most once; the constraint is entirely configurational.
One of the most prominent characteristics of these systems is their fractal geometry,
the ubiquity of which is nowadays a well-established observation \cite{Mandelbrot:1982}.
Mappings exist between ordinary and geometric critical phenomena, such as
the celebrated result of de Gennes \cite{deGennes:1972}
(proved non-perturbatively in \cite{CarvalhoCaracciolo:1983})
concerning the $N\to 0$ limit of the $O(N)$ model.
Geometric phenomena are thus interesting both \emph{per se} and for their descriptive power,
and also as testing grounds for ideas and tools in statistical mechanics.

In this perspective, it is perhaps not surprising that certain properties of percolation
can be mapped \cite{Coniglio:1987} to those of an interacting version of SAW (called ISAW), in which
adjacent occupied sites give a negative contribution to the energy, i.e.\ they attract each other,
thus counteracting the opposite tendency due to the self avoidance.
ISAW are a good model for polymers in solution, as they provide an additional tunable
parameter that can modulate the strength of the effective solvent-mediated interaction between
monomers, on top of the repulsive excluded-volume effect.
As the temperature is varied, the long-chain behavior undergoes a phase transition
between an elongated SAW-like phase and a dense globular one.
The transition, called the $\theta$ point, is tricritical.
In fact, polymeric behavior in two spatial dimensions is diverse, and lattice models
are being invented or rediscovered that display as surprising as rich a phenomenology
\cite{JacobsenKondev:2004,BediniOwczarek:2013}.

It is decisive, in the study of polymers and geometric critical phenomena, to
have available the most assorted set of tools, both theoretical and numerical.
Efforts and results in both areas have often proceeded on the same pathway,
with enriching exchanges in both directions \cite{vanRensburg:2009}.
A new tool often comes in the form of a new model, perhaps easier to analyze
or to simulate, perhaps providing deeper insight \cite{GaroniGuttmann:2009}.
A strategy recently proposed stems from the definition of ``endless''
walks \cite{Clisby:2013}, i.e.~lattice walks that can be infinitely concatenated,
thus excluding any finite-chain effects.
A similar property holds for another ensemble \cite{Gherardi:2010}, based on
a suitable discretization of Schramm-Loewner evolution (SLE),
which gives rise to an exact-sampling algorithm for
self-avoiding paths in the complex plane, in the same universality class of SAW.

Here, with these motivations in mind, we first explore the connection between SLE
and $\theta$-point polymers, showing how the two models are connected.
Then we employ the aforementioned discretized SLE
(such that the Euclidean length of the steps is approximately constant along the curve)
to explore the connection further and to measure critical exponents
and properties related to the shape of the curves.

SLE (for a nice introduction ``for physicists'' see \cite{Cardy:SLEreview})
is a stochastic process taking values on conformal maps in a domain.
Let us fix the domain to be the unit circle $\mathbb{D}$.
For each time $t$, a random map $g_t(z)$
is identified as the solution to the \emph{Loewner equation}
\begin{equation}\label{eq:SLE}
\partial_t g_t(z) = g_t(z)\frac{\exp(\rmi\xi_t)+g_t(z)}{\exp(\rmi\xi_t)-g_t(z)},
\quad g_0(z)=z,
\end{equation}
where the \emph{driving function} $\xi_t$ is the one-dimensional
stochastic process $\xi_t=\sqrt{\kappa}B_t$;
$\kappa$ is a parameter (playing a crucial role in the theory)
and $B_t$ is standard Brownian motion.
The complement of the domain of $g_t$, or equivalently
the set of points for which (\ref{eq:SLE}) does not
admit a solution up to time $t$, is a random fractal curve,
the properties of which depend on $\kappa$; notably,
their fractal dimension is $1+\kappa/8$.
The curves germinate from the point $z=\exp(\rmi\xi_0)$
and grow towards $z=0$ for large times.
This geometry is called radial,
as opposed to the chordal one where the curve
grows from a boundary point to a boundary point.
By complex inversion, one can then define curves living
in the complement of $\mathbb{D}$ in the complex plane,
growing towards $z=\infty$
(see Sec.~\ref{section:drivingfunction}).
SLE has been shown (either rigorously or by numerical and approximate means)
to give the scaling behavior of a large number of critical models
for different values of $\kappa$
\cite{AmorusoHartmann:2006,Boffetta:2006,GliozziRajabpour:2010,Duminil-CopinSmirnov:2011,Daryaei:2012}.
Moreover it is the starting point for many generalisations and extensions
\cite{MoghimiAraghiRajabpourRouhani:2004,RushkinOikonomouKadanoffGruzberg:2006,
Nezhadhaghighi:2010, GherardiNigro:2013}.

SLE is a natural candidate for the continuum counterpart of polymers,
as it describes non-self-crossing critical curves.
Its connection with conformal field theory \cite{BauerBernard:2003}
sets a relation between the central charge $c$ and the parameter $\kappa$,
namely $c=(8-3\kappa)(\kappa-6)/(2\kappa)$.
The only two values corresponding to $c=0$, which is expected
for critical and tricritical polymers, are $\kappa=8/3$,
which is known to describe SAW \cite{LawlerSchrammWerner:2004, Kennedy:2002prl},
and $\kappa=6$, which corresponds to percolation \cite{CamiaNewman:2007}
and is the obvious suspect for the $\theta$ transition.
The fractal dimension, which has been calculated exactly
for the latter \cite{DuplantierSaleur:1987}, agrees with the SLE value $7/4$.

Curves in SLE, as we already noted, naturally extend to infinity,
and considering the evolution up to a time $t$ equates to taking
a finite portion of the whole curve.
Lattice polymers, on the contrary, are affected by
finite-chain effects, due to their not extending to infinity;
these effects are caused by configurations
where the endpoint is ``trapped'' in a loop,
which can appear at every scale at the critical point,
so that finite-chain behavior persists even
when the degree of polymerization $n$ (i.e.~the length of the polymers)
is sent to infinity in the critical limit.
When endpoint distribution functions are considered, for example,
the scaling variable is $\rho=\left|\vect{r}\right|/\xi$,
where $\vect{r}$ is the position of the endpoint and $\xi$
is a correlation length; both $\xi$ and $\vect{r}$ scale as 
some power of $n$ (see Sec.~\ref{section:distributionfunctions}).
Moreover, the match between quantities on the SLE side with
the corresponding properties of lattice polymers requires
consideration of whether they are parameterization-independent
or not, as counting the number of steps on the lattice does not
in general correspond to measuring time in the continuum model.
As far as SAW are concerned, these issues have been
addressed \cite{Gherardi:2009,Gherardi:2010}: the radial distribution
of a properly chosen point along an SLE curve at $\kappa=8/3$
matches that of an \emph{internal} point in a lattice SAW;
in other words, the polymers are ``endless'', in the spirit of \cite{Clisby:2013}.
There are two ways of choosing the right point on the SLE side: first, one can measure
the fractal variation along the curve and stop when it reaches a given value \cite{Kennedy:2007};
second, one can produce discrete curves (chains) so that consecutive points
lie an approximately constant distance apart from each other \cite{Gherardi:2010}.
The latter is the strategy we employ here.

The model of interacting self-avoiding walks that we are going to adopt
is the standard ISAW on the square lattice.
The ensemble is that of all self-avoiding nearest-neighbor $n$-step walks $\omega$,
with the Gibbs measure defined by the following energy function:
\begin{equation}
H(\omega)=-\sum_{i=0}^{n-3}\sum_{j=i+3}^n \delta_{\left|\omega_i-\omega_j\right|,1},
\end{equation}
which simply counts the number of nearest-neighbor contacts between non-bonded monomers.
We will study the model at the 
inverse temperature corresponding to the transition between good and poor solvent, $\beta_\theta$.
The best estimate on the square lattice is $\beta_\theta=0.6673(5)$ \cite{CaraccioloGherardi:2011};
we will use this central value in the following.

\section{Driving function}\label{section:drivingfunction}

An approximation to the driving function can be obtained by the \emph{zipper}
prescription \cite{MarshallRohde:2007}.
Schematically, it amounts to applying iteratively a simple conformal map to all points
of a lattice walk, mapping them in sequence to the boundary.
It is usually done in the half plane, by means of vertical-slit or tilted-slit mappings, 
but we will apply the same strategy on the plane.

Consider the conformal map
\begin{equation}\label{eq:incrementalmap}
\begin{aligned}
\phi_t(z)= \frac{1}{2\rme^t z}&\left[
(z+1)^2-2\rme^t z\right.\\
&-\left.(z+1)\sqrt{(z+1)^2-4\rme^t z}\right],
\end{aligned}
\end{equation}
which is the solution to (\ref{eq:SLE}) for a constant $\xi_t=0$.
The domain which gets mapped by $\phi_t$ onto the
unit disc is the unit disc with a radial slit removed;
the slit sticks out of $z=1$ and is directed towards $z=0$.
The length of the slit increases with $t$ non-linearly.
Let us consider the map
\begin{equation}\label{eq:inverseincrementalmap}
f_t(z)=\frac{1}{g_t(1/z)};
\end{equation}
it sends the exterior of the slitted disc conformally onto the exterior of the disc.
The expansion around infinity of its inverse,
\begin{equation}
f^{-1}_t(w)=\rme^t w+2\left(\rme^t-1\right)+\mathcal{O}(1/w),
\end{equation}
identifies the \emph{(logarithmic) capacity} of the slitted disc $\mathrm{cap}(t)=\rme^t$,
and its \emph{conformal center} $c(t)=2\left(\rme^t-1\right)$.

Let $\left\{\omega_i\right\}$, $i=0,\ldots,n$, be a lattice walk.
Here, the lattice is embedded in the complex plane (say, with lattice spacing $a=1$).
The zipper algorithm is a prescription to find a driving function
giving rise (via the Loewner equation) to an approximation of $\omega$.
Operatively, at the $k$-th step one applies, to all points 
$\left\{\omega_i\right\}$ with $i\geq k$,
the radial-slit mapping $\varphi$ that maps $\omega_k$ to the unit circle, that is
\begin{equation}
\varphi\left(\omega_k;z\right)= f_{t_k}\left(\rme^{-\rmi\theta_k} z\right),
\end{equation}
where
\begin{equation}
\theta_k=\arg\left(\omega_k\right)
\end{equation}
and the time $t_k$ is obtained by solving $f_{t_k}\left(\left|\omega_k\right|\right)=1$, i.e.,
\begin{equation}
t_k=\log\frac{\left(\left|\omega_k\right|+1\right)^2}{4\left|\omega_k\right|}.
\end{equation}
Now, $\theta_k$ is a compact variable, but since the driving function for SLE
is supposed to be continuous then the interval $[0,2\pi]$ can be
unwrapped univocally onto the whole real line to obtain the approximate
driving function $\xi_k$.
Operatively, we set $\xi_0=\theta_0=w_0=0$ and compute $\delta_k=\theta_k-\theta_{k-1}$
at each step; whenever $\left|\delta_k\right|>\pi$ we increase or decrease
the winding number $w_k$. The driving function is then $\xi_k=\theta_k+2\pi w_k$.

\begin{figure}
\centering
\includegraphics[scale=1]{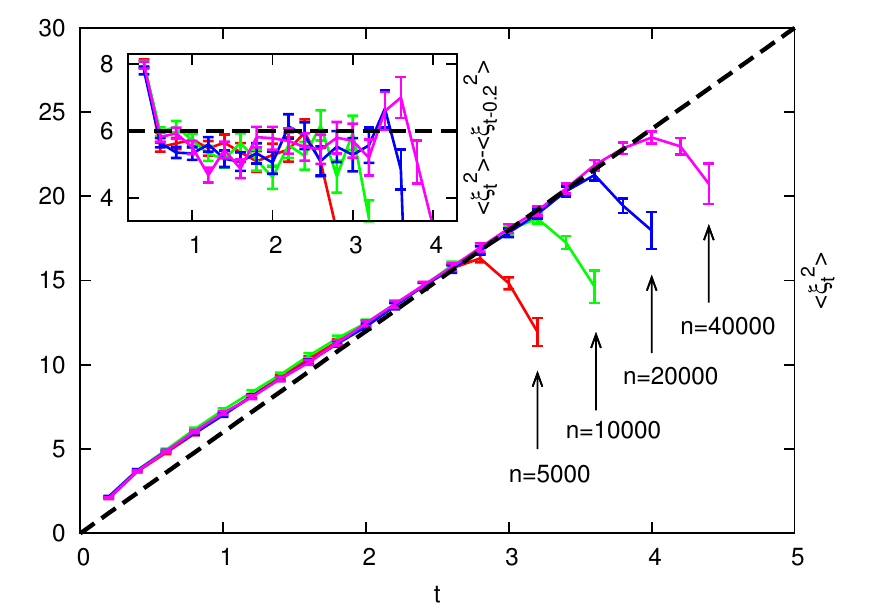}
\caption{
(Color online)
After an initial transient, the mean square displacement of the driving function
is linear in time, as expected for Brownian motion
(time corresponds to the usual parameterization by capacity, see text).
The slope is compatible with $\kappa=6$ (dashed line).
The four curves correspond to walks of different lengths;
in each case only the first $1/10$ of the steps were used
in the analysis.
The inset shows the discrete derivative of $\left<\xi_t^2\right>$
computed from the same data.
}
\label{figure:kappa}
\end{figure}

We generated $\theta$-point walks in the full plane with increasing lengths
($n=5000,10\,000,20\,000,40\,000$) and obtained $10\,000$
samples for each $n$.
The method used is the extended reptation algorithm \cite{CaraccioloCauso:2000,CaraccioloPapinutto:2002};
remark that very efficient algorithms are available for the non-interacting SAW \cite{Clisby:2010,Kennedy:2002},
but they are expected to perform much less efficiently for the ISAW.
In order to reduce the severity of finite-chain effects, we restricted the analysis
of the driving function to the first $n/10$ steps (we obtain similar results
with $n/20$ steps;
an independent check that corrections in this regime are small
is given in Fig.~\ref{figure:distributions} for the radial distribution functions,
see Sec.~\ref{section:distributionfunctions}).
The function $\xi_t$ computed by the zipper algorithm turns out to be
compatible with a Brownian motion with the expected diffusivity $\kappa=6$,
after an initial transient where deviations from this behavior dominate.
The plots in Fig.~\ref{figure:kappa} show how $\left<\xi_t^2\right>$ depends
on $t$ (the average is over all samples generated).
The expected behavior $\left<\xi_t^2\right>\sim \kappa t$ is recovered
for large times, increasingly well in $n$.
The initial transient displays an effectively higher diffusivity
(around the critical value $\kappa=8$, where curves become space-filling) 
followed by a lower one, which slowly
relaxes towards $\kappa=6$.
We note that in the whole plane (contrary to the half-plane case)
equal-length steps in the walk contribute a smaller and smaller
capacity as they get further away from the origin.
Therefore, although the initial transient is present up to times $t\approx 2$,
it really regards only the first few walk steps (approximately $100$).
The falloffs at large $t$ are due to the fact that not all generated
walks reach such large capacity; those that do are then biased
to be more straight, and thus their driving function fluctuates less.
Table~\ref{table:kappa} shows numerical values of $\kappa$ obtained
by linear fits with several cutoffs;
the upper cutoffs are chosen so as to exclude the aforementioned biased region.
\begin{figure}[t]
\centering
\includegraphics[scale=1]{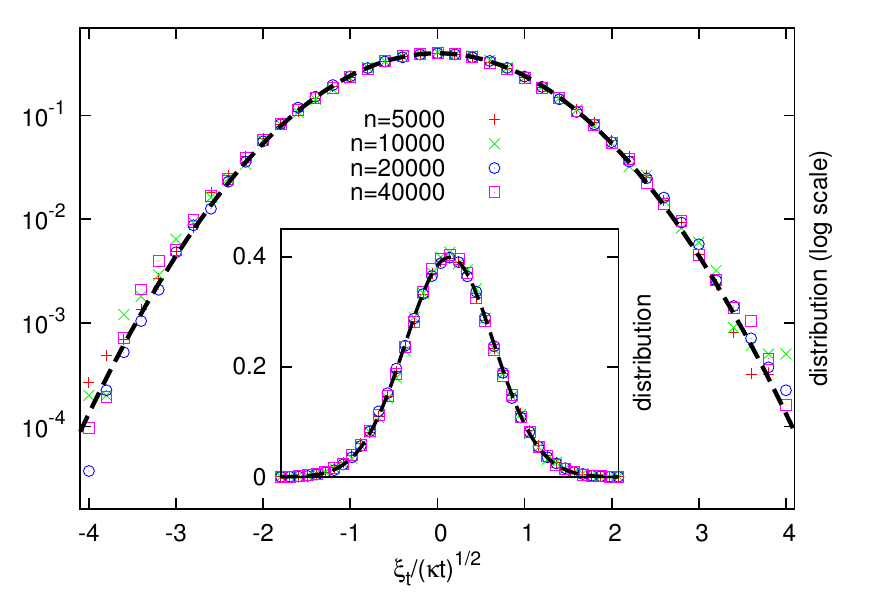}
\caption{
(Color online)
The driving function is Gaussian.
Symbols are histograms of the rescaled variable $\xi_t/\sqrt{\kappa t}$,
where we have fixed $\kappa=6$.
As in Fig.~\ref{figure:kappa}, only the first tenth of the walk was considered.
Moreover, for each $n$, we used only data with $2<t<t_\mathrm{max}(n)$,
where $t_\mathrm{max}(n)$ are the cutoffs in Tab.~\ref{table:kappa}.
}
\label{figure:gaussian}
\end{figure}
\begin{table}[b]
\centering
\begin{tabular}{|r|c|c|c|}
\hline
$n$ & $t_\mathrm{min}$ & $t_\mathrm{max}$ & $\kappa$\\
\hline
5\,000 & 0 & 2.7 & 5.8(1)\\
10\,000 & 0 & 3.1 & 5.7(1)\\
20\,000 & 0 & 3.5 & 5.6(1)\\
40\,000 & 0 & 3.9 & 5.7(1)\\
\hline
5\,000 & 2 & 2.7 & 5.6(1)\\
10\,000 & 2 & 3.1 & 5.5(1)\\
20\,000 & 2 & 3.5 & 5.6(1)\\
40\,000 & 2 & 3.9 & 5.8(1)\\
\hline
20\,000 & 2.5 & 3.5 & 5.67(14)\\
40\,000 & 2.5 & 3.9 & 5.92(10)\\
\hline
\end{tabular}
\caption{
Fit results for the curves in Fig.~\ref{figure:kappa},
with several lower cutoffs on time;
the upper cutoffs are fixed by simple considerations
related to biasing (see the text).
The closest value to the expected $\kappa=6$
is obtained for the longest walks and the
highest cutoff.
}
\label{table:kappa}
\end{table}

If $\xi_t$ is Brownian motion, fluctuations with respect to the average linear
behavior must be normally distributed.
Taking into account the diffusivity, which we will fix to $\kappa=6$, we expect
\begin{equation}
\frac{\xi_t}{\sqrt{\kappa t}}\sim \mathcal{N}(0,1),
\end{equation}
where $\mathcal{N}(0,1)$ is the normal distribution of mean $0$ and variance $1$.
Since the foregoing analysis shows that deviations from the expected
diffusivity are present for $t\apprle 2$ we compute
$\xi_t/\sqrt{\kappa t}$ by using only data with $t>2$.
Figure~\ref{figure:gaussian} shows the results, superimposed on the
Gaussian $p(x)=\exp\left(-x^2/2\right)/(2\pi)^{1/2}$.
As a quantitative test, we compute the Pearson's $\chi^2$ \mbox{$p$-values}
for several cutoffs in $t$ (Tab.~\ref{table:chisquare});
after the transient, the driving function is solidly Gaussian.
\begin{table}[t]
\centering
\begin{tabular}{|r|c|c|c|}
\hline
$n$ & $t_\mathrm{min}$ & $t_\mathrm{max}$ & $p$-value\\
\hline
5\,000 & 2 & 2.7 & 0.67\\
10\,000 & 2 & 3.1 & 0.001\\
20\,000 & 2 & 3.5 & 0.40\\
40\,000 & 2 & 3.9 & 0.004\\
\hline
5\,000 & 2.5 & 2.7 & 0.90\\
10\,000 & 2.5 & 3.1 & 0.12\\
20\,000 & 2.5 & 3.5 & 0.60\\
40\,000 & 2.5 & 3.9 & 0.06\\
\hline
\end{tabular}
\caption{
The goodness-of-fit $p$-value  obtained by the standard
$\chi^2$ test.
All $p$-values for the higher cutoff $t_\mathrm{min}=2.5$
are above the stringent significance level $0.05$.
}
\label{table:chisquare}
\end{table}

\section{Distribution functions}\label{section:distributionfunctions}

The opposite procedure to that of the previous section
can be performed, in such a way as to generate curves $\gamma$ in the complex plane.
Instead of absorbing the points of the walk into the disc by use of 
the incremental map $f_t$, one uses $f_t^{-1}$ to grow them.
The driving function here is fixed to (a suitable discretization of)
Brownian motion, and the collection of points $\left\{\gamma_i\right\}$
is what the algorithm produces.
As a prescription, we approximate $B_t$ by a
piecewise constant function, because then $\phi$ is the solution to (\ref{eq:SLE})
in each one of the intervals where the driving function is constant, as already noted.
Hence, composition of the maps $f_{\delta t}^{-1}$, where $\delta t$ is the time step,
intertwined with rotations by $\theta$, where $\theta$ follows a random walk,
iteratively grows the set $\left\{\gamma_i\right\}$.
This is the standard way that (radial) SLE is simulated, through the
so-called \emph{backward} evolution; for details, see \cite{Gherardi:2009,Kennedy:2009review}.

We need a way of directly comparing distribution functions
of lattice $\theta$-point walks and paths obtained by such a discretization of (\ref{eq:SLE}). 
In order to do so, the \emph{parameterizations} in the two models must coincide.
Schramm-Loewner paths come with a natural proper time, which is the one
imposed by an exponentially increasing logarithmic capacity
(a linearly increasing half-plane capacity in the chordal geometry),
and lattice walks have their own, induced by the requirement
that consecutive points belong to nearest neighbors.
So what we need is a way of generating continuum walks $\gamma$
(on the SLE side) such that the Euclidean lengths of their steps
$\left|\gamma_{i+1}-\gamma_i\right|$ be constant along the chain, at least approximately.
Such a task encounters some technical difficulties,
which nonetheless can be overcome.
The method we employ, first proposed in \cite{Gherardi:2010},
rescales each step in the iterated-map approach
by an appropriate factor, obtained by tracking the evolution
of the Jacobian of the map in an affordable way.
We do not describe the details here, and the reader is referred
to the original article.

\begin{figure}[t]
\centering
\includegraphics[scale=1]{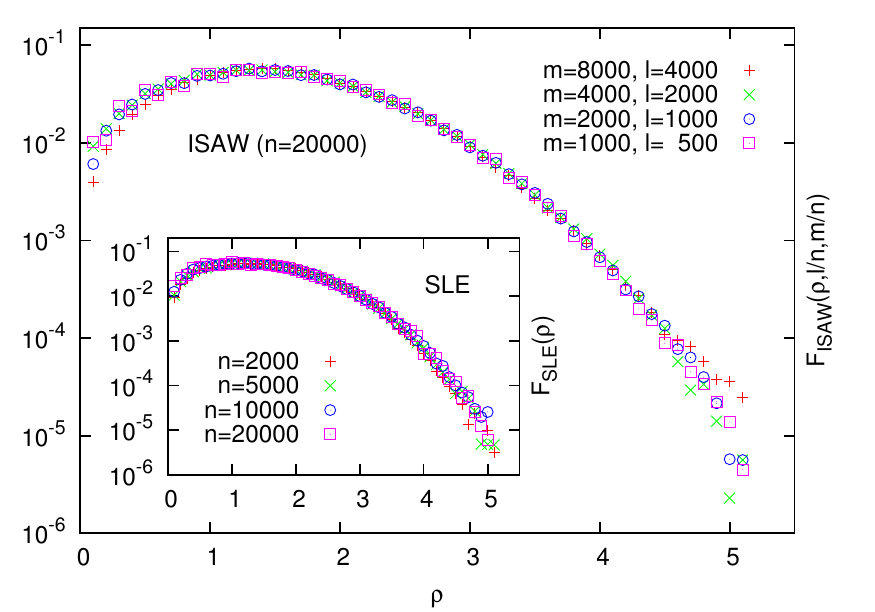}
\caption{
(Color online)
The radial distribution functions of lattice $\theta$-point walks (main panel)
and of discrete SLE chains (inset)
satisfy the scaling forms (\ref{eq:scaling_ISAW})
and (\ref{eq:scaling_SLE}) respectively.
Symbols correspond to different values of the length $n$
for SLE chains, and to different internal points $m$ and $l$ for
lattice walks.
The only detectable deviations belong to the higher values of
$\lambda_1=l/n$ and $\lambda_2=m/n$
(red crosses, with $\lambda_1=0.2$, $\lambda_2=0.4$).
}
\label{figure:distributions}
\end{figure}

Another obstruction 
to the program of matching the distribution functions
of SLE and tricritical polymers is related to the fact that the full plane and
the exterior of the disc are not conformally equivalent.
When dealing with self-avoiding walks (corresponding to $\kappa=8/3$)
one can take the position that conformal restriction, a special property concerning
restrictions of the domain, which is valid only for that value of $\kappa$,
makes the distinction irrelevant.
In fact, no problem arises for the SAW \cite{Gherardi:2009}.
But in the case of $\theta$-point polymers, we are forced
to take care of the domain.
Given a lattice walk $\omega$ of length $n$,
the sub-walk $\hat\omega=\left\{\omega_i\right\}$, $i=0,\ldots,l$, must
be mapped to the unit disc by a conformal map $\phi_{\hat\omega}$ 
(more precisely, 
the complement in $\mathbb{C}$ of the piece-wise linear curve
having vertices in $\left\{\omega_i\right\}$
is mapped onto the exterior of the unit disc).
Then the distribution functions of $\left\{\phi_{\hat\omega}\left(\omega_i\right)\right\}$, 
$i=l+1,\ldots,m$, are expected to match those obtained from SLE,
provided the scaling limit $l,m,n\to\infty$ is performed, with $m\ll n$
(so as to probe the walk far from the endpoint)
and $(m-l)\to\infty$ (scaling limit of the subwalk).
Note that this entails the evaluation of a random (i.e., realization dependent)
conformal map for each walk.

We do not embark here on such an exploration,
and rather pursue an approximation.
Namely, we consider the distribution functions for the internal
sub-walk $\left\{\omega_i\right\}$, $i=l+1,\ldots,m$ itself,
without applying the map $\phi_{\hat\omega}$.
We will suppose that, given an $n$-step $\theta$-point walk $\omega$,
the probability that the $l$-th and $m$-th points (with $l<m$) are such that 
$\omega_m-\omega_l=\vect{r}$
has the following scaling form, when $n,m,l\to\infty$, $\vect{r}\to\infty$, 
keeping $l/n=\lambda_1$ and $m/n=\lambda_2$ fixed:
\begin{equation}\label{eq:scaling_ISAW}
P_{n,l,m}(\vect{r})\sim\frac{1}{\xi_{l,m}^2}F_\mathrm{ISAW}\left(\rho,\lambda_1,\lambda_2\right),
\end{equation}
where $\rho=\left|\vect{r}\right|/\xi_{l,m}$ and $\xi_{l,m}^2=\left<\left|\omega_m-\omega_l\right|^2\right>$.
(In measuring $F_\mathrm{ISAW}$ from data, lattice effects have been alleviated
by a suitable averaging procedure that takes into account
the number of lattice points in each annulus $[\rho,\rho+\Delta\rho]$.)
After taking this continuum limit, we want to let $\lambda_2\to 0$
(and hence also $\lambda_1\to 0$),
in order to exclude the finite-chain effects due to the presence of the endpoint.
Such a complicated non-commutative limit is tricky to analyze numerically,
so we give up a detailed quantitative study, and just show that
the distribution functions for fixed $n$ and several values of $l$ and $m$,
calculated as in (\ref{eq:scaling_ISAW}), collapse on the same curve,
thus confirming the scaling form (see Fig.~\ref{figure:distributions};
deviations from the scaling behavior are apparent only for the
largest values of $m$ and $l$, i.e., when $\lambda_1$ and $\lambda_2$
are significantly different from $0$).
On the SLE side,
the distribution function is defined on the same lines.
The probability density that the $n$-th point along the discrete chain $\gamma$
is in $z$, for $n\to\infty$ and $z\to\infty$, behaves as
\begin{equation}\label{eq:scaling_SLE}
P_n(z)\sim\frac{1}{\xi_n^2}F_\mathrm{SLE}\left(\rho\right),
\end{equation}
where $\rho=zz^*/\xi_n$ and $\xi_n^2=\left<\gamma_n\gamma_n^*\right>$.
This scaling form has been already verified numerically for $\kappa=8/3$ \cite{Gherardi:2009},
and here we check it for $\kappa=6$;
the inset in Fig.~\ref{figure:distributions} shows the collapse of $F_\mathrm{SLE}(\rho)$
on the universal curve for chains of different lengths.

\begin{figure}[t]
\centering
\includegraphics[scale=1]{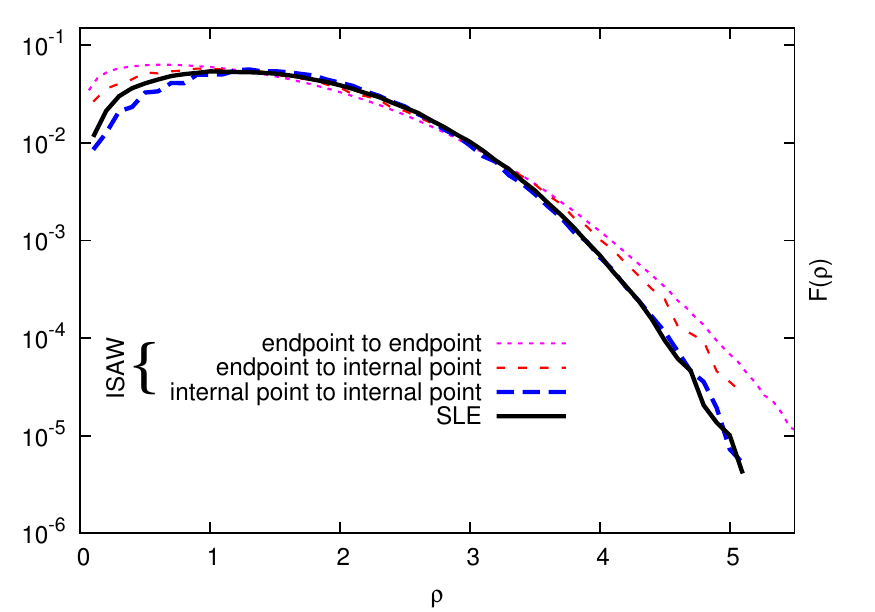}
\caption{
(Color online)
The radial distribution functions of the endpoint of an SLE chain
and of the vector connecting two internal points deep inside
an ISAW at the $\theta$-point agree in the large-$\rho$ regime
and slightly disagree for $\rho\apprle 3$
(thick black solid line and thick blue dashed line respectively).
For comparison, two other distribution functions for ISAW
are shown with thinner lines, namely endpoint-to-endpoint
(pink dotted line) and internal-to-endpoint (red dashed line).
}
\label{figure:distributions2}
\end{figure}

Finally, we compare the distribution functions $F_\mathrm{SLE}(\rho)$ and
$F_\mathrm{ISAW}(\rho,\lambda_1\ll 1,\lambda_2<\lambda_1)$
by collapsing all data with $n=5000,10\,000,20\,000$ for SLE
and all data with $m=1000,2000,4000$ for ISAW.
As argued at the beginning of this section, we do not expect the two
functions to be equal. Nonetheless, effects due to the domain
are supposed to be be irrelevant in the large-$\rho$ regime,
so $F_\mathrm{ISAW}$ and $F_\mathrm{SLE}$ should agree
in this limit.
Figure~\ref{figure:distributions2} shows qualitatively that this is the case,
and additionally that the distribution functions only slightly disagree
for small $\rho$.
Fits against the ansatz $\log F(\rho)=c - D\rho^\delta$,
inspired by the known large-distance form for $\theta$-point ISAW \cite{CaraccioloGherardi:2011},
yield $\delta=3.2(3)$ in the case of $F_\mathrm{ISAW}$
(by averaging over windows with different lower cutoffs on $\rho$),
and $\delta=3.1(1)$ in the case of $F_\mathrm{SLE}$;
values of $c$ and $D$ are compatible as well.
In contrast, the endpoint distribution function and that for a single
internal point, corresponding to setting $l=0$ in (\ref{eq:scaling_ISAW}),
have very different shapes (Fig.~\ref{figure:distributions2}).

\section{Critical exponents and the gyration tensor}

A useful quantity in the study of polymers is the
\emph{end-to-end} distance (or \emph{radius}), i.e., the distance from the origin
to the endpoint, which measures the elongation of the walk.
Here, we consider the radius for the generic internal point $m$ in an SLE chain $\gamma$,
which has the same definition as the correlation length, $R^2_m=\gamma_m\gamma_m^*$.
In our framework $R^2_m$ measures the elongation of an
endless $\theta$-point polymer of length $m$.
This quantity has the usual structure of corrections to scaling:
\begin{equation}\label{eq:correctionstoscaling}
\frac{\left<R^2_m\right>}{m^{2\nu}}=
a+\frac{a_1}{m}+\frac{a_2}{m^2}+\cdots+\frac{b_0}{m^{\Delta_1}}+\frac{b_1}{m^{\Delta_1+1}}+\cdots
\end{equation}
where the first terms are analytical corrections
and the others have non-integer exponents.
The asymptotic form (\ref{eq:correctionstoscaling}) defines the Flory exponent $\nu$
and the leading correction-to-scaling exponent $\Delta_1$.
The numerical estimation of $\Delta_1$ is often complicated by
the superposition of different terms (especially if they have
discording signs \cite{CaraccioloGuttmann:2005}) and by lattice artifacts.
In the case of self-avoiding walks the SLE approach has proven
interesting \cite{Gherardi:2010}, as it displays the corrections with $\Delta_1=11/16$
that are predicted by conformal field theory \cite{Saleur:1987} but not detectable
for lattice self-avoiding walks \cite{CaraccioloGuttmann:2005}
(not even in their endless formulation \cite{Clisby:2013}).
Let us fix an increment $\delta m$ and consider $\left<R^2_m\right>$
when $m$ varies in steps of length $\delta m$.
We can define an effective Flory exponent $\nu_\mathrm{eff}$ as
\begin{equation}\label{eq:effectiveexponent}
\nu_\mathrm{eff}(m)=\frac{1}{2} \log \frac{\left<R^2_{m+\delta m}\right>}{\left<R^2_m\right>}
\left(\log \frac{m+\delta m}{m}\right)^{-1},
\end{equation}
which is simply equal to the critical exponent $\nu$ if the end-to-end radius
is exactly proportional to $m^{2\nu}$.
Its dependence on $m$ will highlight the corrections to scaling;
in particular, if only the term with exponent $\Delta_1$ 
in (\ref{eq:correctionstoscaling}) is kept, one has 
$\nu_\mathrm{eff}(m)-\nu\propto m^{-\Delta}$ asymptotically.

\begin{figure}[t]
\centering
\includegraphics[scale=1]{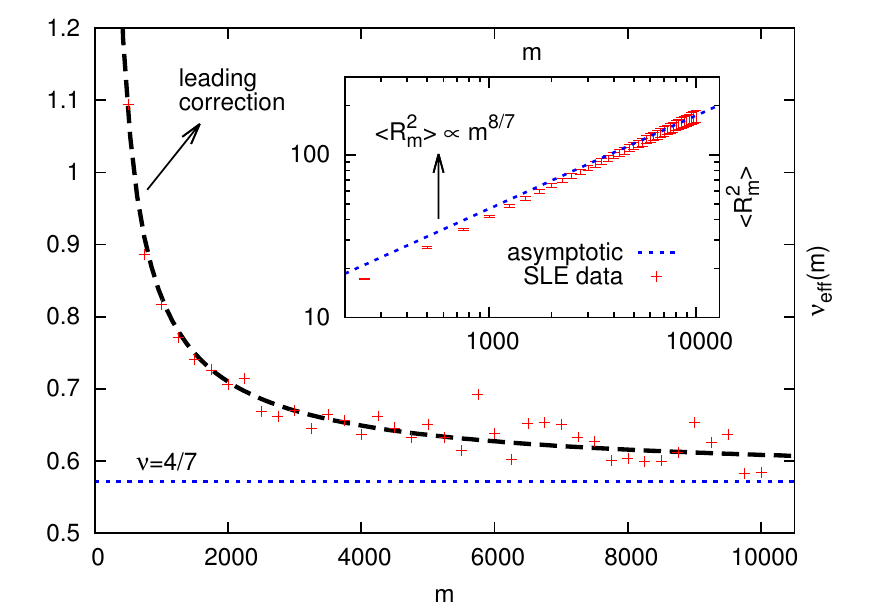}
\caption{
(Color online)
The asymptotic behavior of the end-to-end radius for SLE chains
is consistent with that of $\theta$-point polymers.
The effective exponent $\nu_\mathrm{eff}$ measured from simulations
(red crosses) is plotted against the chain length $m$; the dotted blue line
is the asymptotic value corresponding to the $\theta$-point prediction;
the dashed black line includes the fitted corrections to scaling.
The inset shows the end-to-end radius, from which the effective
exponent is computed, together with its asymptotic expression.
}
\label{figure:radius}
\end{figure}

We generated $\sim 150\,000$ independent samples of
discrete SLE chains of length $n=10\,000$ (at $\kappa=6$), and measured
$\left<R^2_m\right>$ for $m=j\cdot\delta m$, with $j=1,\ldots,40$ and $\delta m = 250$.
Notice that the values obtained for different values of $m$ are not independent,
so our estimates are probably affected by slight
systematic errors that are difficult to quantify
(the uncertainties are probably underestimated as well).
Setting the ansatz $\left<R^2_m\right>\propto m^{2\nu}(c + m^{-\Delta_1})$
in (\ref{eq:effectiveexponent}) gives a form against which data can be compared.
Varying both exponents $\nu$ and $\Delta_1$ in the fit is unfeasible,
but Fig.~\ref{figure:radius} shows that
the effective exponent converges to the expected value $\nu=4/7\approx 0.5714$.
This is confirmed by fits of the form $\left<R^2_m\right>\propto m^{2\nu}$
for the radius, performed in windows $m\in[m_\mathrm{min},10\,000]$ for
increasing values of the cutoff,
which give $\nu=0.60$ for $m_\mathrm{min}=6000$,
$\nu=0.59$ for $m_\mathrm{min}=7000$ and $8000$,
and $\nu=0.58$ for $m_\mathrm{min}=9000$.
Hence we fix $\nu$ to its theoretical value and fit the correction-to-scaling
exponent to all available data
(setting a lower cutoff increases sensibly the errors,
but the values are all compatible with the one given below).
The results of this procedure (plotted in Fig.~\ref{figure:radius}) bring to
\begin{equation}\label{eq:delta1}
\Delta_1 = 0.76(5).
\end{equation}
This exponent, to our knowledge, has never been computed for $\theta$-point polymers,
and its value in percolation has been the subject of long debate \cite{Ziff:2011talk}.
Our result (\ref{eq:delta1}) is in perfect agreement with the latest estimates 
for percolation and with the theoretical prediction $\Delta_1=72/91\approx 0.791$
obtained independently from calculations in the Potts model \cite{AharonyAsikainen:2003}
and in the $O(n)$ model \cite{Ziff:2011}.
However, caution must be exercised;
as already noted,
corrections to scaling are a notoriously slippery ground,
since terms with different exponents (and possibly different signs) 
can conspire to yield deceptive effective exponents.
Further confirmations of this result, perhaps obtained with other techniques, are awaited.

Although rotational symmetry forces the distribution functions to be spherically
symmetric, the instantaneous shape of polymers is not spherical on average.
A measure of the deviations from spherical shape is given by the \emph{gyration tensor}
\begin{equation}
Q_{\alpha\beta}=\frac{1}{2(n+1)^2}\sum_{i,j=0}^n
\left(\gamma_{i,\alpha}-\gamma_{j,\alpha}\right)\left(\gamma_{i,\beta}-\gamma_{j,\beta}\right),
\end{equation}
where $\gamma_{i,\alpha}$ is either $\mathrm{Re}(\gamma_i)$ or $\mathrm{Im}(\gamma_i)$
depending on the value of $\alpha=0,1$
(note that it implicitly depends on $n$).
$Q_{\alpha\beta}$ is symmetric and positive definite;
therefore it has two positive eigenvalues, $q_1$ and $q_2$, say such that $q1\geq q2$.
Two quantities characterizing the shape of the curves can be constructed from the eigenvalues:
\begin{equation}
r_n=\frac{\left<q_1\right>_n}{\left<q_2\right>_n}, \quad
\mathcal{A}_n=\left<\frac{\left(q_1-q_2\right)^2}{\left(q_1+q_2\right)^2}\right>_{\!\!n},
\end{equation}
called \emph{shape factor} and \emph{asphericity} respectively.
For a spherically symmetric object, $r=1$ and $\mathcal{A}=0$,
while in the extreme case of a rod-like curve the asphericity is $1$
and the shape factor diverges.
These quantities are expected to have
a finite limit for large $n$, and behave as $r_n\sim r_\mathrm{SLE} - a/n^{\Delta_1}$
and $\mathcal{A}_n\sim \mathcal{A}_\mathrm{SLE} - b/n^{\Delta_1}$
to leading order in $n$.

\begin{figure}[t]
\centering
\includegraphics[scale=1]{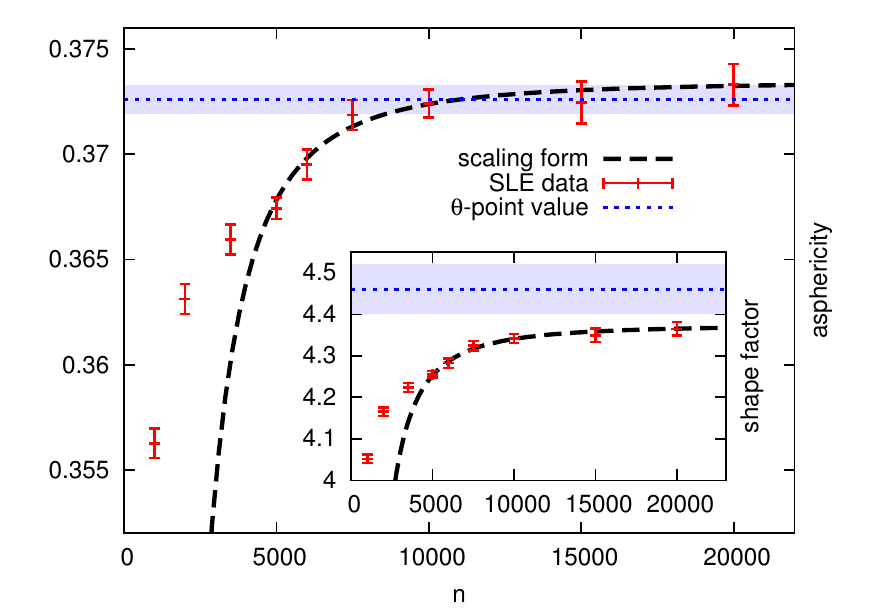}
\caption{
(Color online)
The shape-related characteristics of the discrete SLE curves are surprisingly
close to their $\theta$-point values.
Red symbols are the results of simulations, the black dashed lines
are leading-order scaling forms fitted against data points,
the shaded regions are the confidence intervals for the
best known $\theta$-point estimates (blue dotted lines).
}
\label{figure:gyration}
\end{figure}

We analyzed the shape factor and the asphericity for the SLE chains,
by performing independent simulations at 
$n=1000$, $2000$, $3500$, $5000$, $6000$, $7500$, $8500$, $10\,000$, $15\,000$, $20\,000$.
We remind that the chains are generated in such a way as to have constant step lengths.
Approximately $100\,000$ samples were produced for each $n$,
except $n=5000, 10\,000$ ($200\,000$ samples) 
and $n=15\,000,20\,000$ (about $50\,000$ samples).
Results are presented in Fig.~\ref{figure:gyration}.
Unexpectedly, we find that the predicted scaling form is violated
considerably for $n\lesssim 5000$. Therefore we fix a lower cutoff
and fit the results only for $n>5000$.
From the fits we obtain
\begin{equation}
r_\mathrm{SLE}=4.38(3), \quad \mathcal{A}_\mathrm{SLE}=0.373(1),
\end{equation}
which shows that the curves are strongly elliptical.
The estimated values of $\Delta_1$ are affected by huge errors
($\Delta_1=1.8\pm 1.0$ and $2.4\pm 1.6$, for $r$ and $\mathcal{A}$ respectively)
and are thus not comparable to theoretical predictions,
but they are compatible with the more precise estimate obtained above
from the end-to-end radius
(repeating the fits by keeping $\Delta_1$ fixed to an extreme of its confidence interval
yields $r_\mathrm{SLE}=4.33\sim 4.42$ and $\mathcal{A}_\mathrm{SLE}=0.370\sim0.377$).
The asymptotic values are surprisingly close to those of ordinary
$\theta$-point polymers in two dimensions; in fact, they are
compatible within statistical significance.
The best estimates for polymers,
depicted by the shaded intervals in Fig.~\ref{figure:gyration}, 
are $r_\theta=4.46(6)$ and $\mathcal{A}_\theta=0.3726(7)$ \cite{CaraccioloGherardi:2011}.
By comparison, self-avoiding walks in the plane and discrete SLE
curves with constant step lengths at $\kappa=8/3$ have 
$\mathcal{A}_\mathrm{SAW}\approx 0.5134$ \cite{Gherardi:2010},
while for true random walks $\mathcal{A}_\mathrm{RW}\approx 0.3964$ \cite{Diehl:1989}.

We do not have an explanation for the coincidence of the asphericity measures.
As already noted, the influence of the finite-chain effects and of the domain
(the plane punctured by a disc) modifies the shape and the distribution functions
(Fig.~\ref{figure:distributions2}).
The asphericity for SAW has already been shown to have a non-trivial dependence
on the position of the internal point \cite{Gherardi:2010}.
Therefore the similarity in the overall shape that we observe for $\kappa=6$
and the $\theta$ point is intriguing.
Our result for the exponent $\Delta_1$, instead, fits into the theoretical picture,
and shows the power of the numerical method.
As already noted, the SLE-based numerical strategy
applied to the case $\kappa=8/3$ \cite{Gherardi:2010}
exposed, for self-avoiding paths, the
presence of leading corrections to scaling with the
exponent predicted by conformal field theory, $11/16$,
which are surprisingly absent for square-lattice SAW \cite{CaraccioloGuttmann:2005}
(where the leading corrections have exponent $3/2$).
In this perspective, it would be interesting to calculate $\Delta_1$ directly for $\theta$-point ISAW, and
also for their endless counterpart on the lattice,
as was done in \cite{Clisby:2013} for non-interacting SAW.

\begin{acknowledgments}
I am grateful to Nathan Clisby, Andrea Bedini and Sergio Caracciolo 
for stimulating discussions and for a critical reading of the manuscript,
and to Andrea Sportiello for valuable suggestions.
Computer time and facilities were made available by
Alessandro Vicini and Universit\`a degli Studi di Milano.
This work was supported by Fondo Sociale Europeo
(Regione Lombardia) through the grant ``Dote Ricerca.''
\end{acknowledgments}

\vfill

\bibliographystyle{unsrt}
\bibliography{/Users/marcogherardi/Lavoro/bibliografie/SLE,/Users/marcogherardi/Lavoro/bibliografie/Polymers}

\end{document}